\documentclass[prb,twocolumn,showpacs,floatfix]{revtex4}
\usepackage{graphicx}% Include figure files
\usepackage{dcolumn}% Align table columns on decimal point
\usepackage{bm}% bold math
\begin{document}
\title{Magnetoelastic effects in multiferroic HoMnO$_3$}
\author{Tapan Chatterji$^1$, Thomas Hansen$^1$, Simon A.J. Kimber$^2$  and Dipten Bhattacharya$^3$}
\address{$^1$Institut Laue-Langevin, B.P. 156, 38042 Grenoble Cedex 9, France\\
$^2$European Synchrotron Radiation Facility, B.P. 220, 38043 Grenoble Cedex 9, France\\
%$^3$Helmholtz Zentrum Berlin, Hahn-Meitner-Platz 1, 14109 Berlin, Germany\\
$^3$Central Glass and Ceramic Research Institute, Kolkata 700032, India
}
\date{\today}
\begin{abstract}
We have investigated magnetoelastic effects in multiferroic HoMnO$_3$ below the antiferromagnetic phase transition $T_N \approx 75$ K by neutron powder diffraction. The lattice parameter $a$ of the hexagonal unit cell of HoMnO$_3$ decrease in the usual way at lower temperatures and then shows abrupt contraction below $T_N$ whereas the lattice parameter $c$ increases continuously with decreasing temperature and shows an anomalous increase below $T_N$.  The unit cell volume decreases continuously with decreasing temperature and undergoes abrupt contraction below $T_N$. By fitting the background thermal expansion for a nonmagnetic lattice with the Einstein-Gr\"uneisen equation we determined the lattice strains $\Delta a$, $\Delta c$ and $\Delta V$ due to the magnetoelastic effects as a function of temperature.  We have also determined the temperature variation of the ordered magnetic moment of Mn ion by fitting the measured Bragg intensities of the nuclear and magnetic reflections with the known crystal and magnetic structure models. 
\end{abstract}
\maketitle
One of the most challenging subjects of the condensed matter physics is the coupling between spin and lattice degrees of freedom.  The spin system in ferromagnetic or antiferromagnetic crystals is coupled to the ionic displacements via the dependence on distance of the exchange interaction or spin-orbit or dipole-dipole interaction. The coupling  is called the magnetoelastic coupling and and resultant effects are called magnetoelastic effects \cite{becker39,callen63,callen65}. The static part of this effect is the shift in the equilibrium ionic positions with resultant displacements of the phonon and magnon spectra. The dynamic part of this interaction is the magnon-phonon scattering and also their hybridization. The simplest static part of the interaction is the external magnetostriction, or the change in macroscopic crystal dimensions. This changes the lattice parameters below the magnetic phase transition and can be more easily determined experimentally. In addition there are shifts in the atomic cooodinates within each unit cell. This is called internal magnetostriction which is more difficult to measure. The induced atomic displacements may or may not modify the symmetry of the lattice and also the anisotropic energy. There exist a large number of review articles that summarize the experimental results and also to some extent some theoretical calculations \cite{lee55,clark80,andreev95,lindbaum02,doerr05}. The magnetoelastic effect 
has drawn renewed interest in connection with the potentially useful colossal magnetoresistive and multiferroic materials. Here we have investigated the magnetostriction in multiferroic hexagonal manganite HoMnO$_3$.

\begin{figure}
\resizebox{0.45\textwidth}{!}{\includegraphics{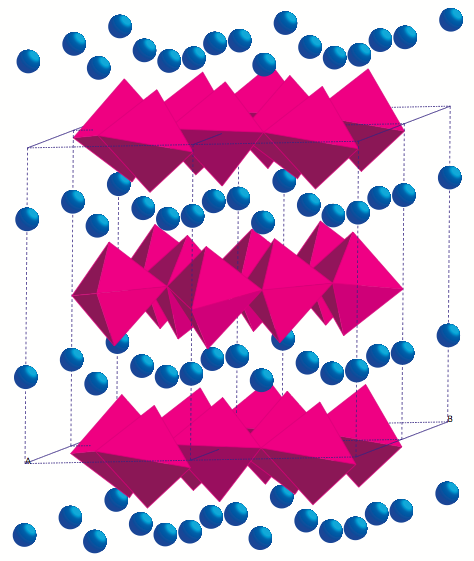}}
\caption {(Color online) Schematic representation of the hexagonal crystal structure of RMnO$_3$ (R = Sc, Y, Er, Ho, Tm, Yb, Lu). Here we have used the structural parameters for HoMnO$_3$. The red trigonal bipyramids  represent Mn ions surrounded by five O atoms and the blue spheres are Ho ions.  }
\label{structure}
\end{figure}
\begin{figure}
\resizebox{0.5\textwidth}{!}{\includegraphics{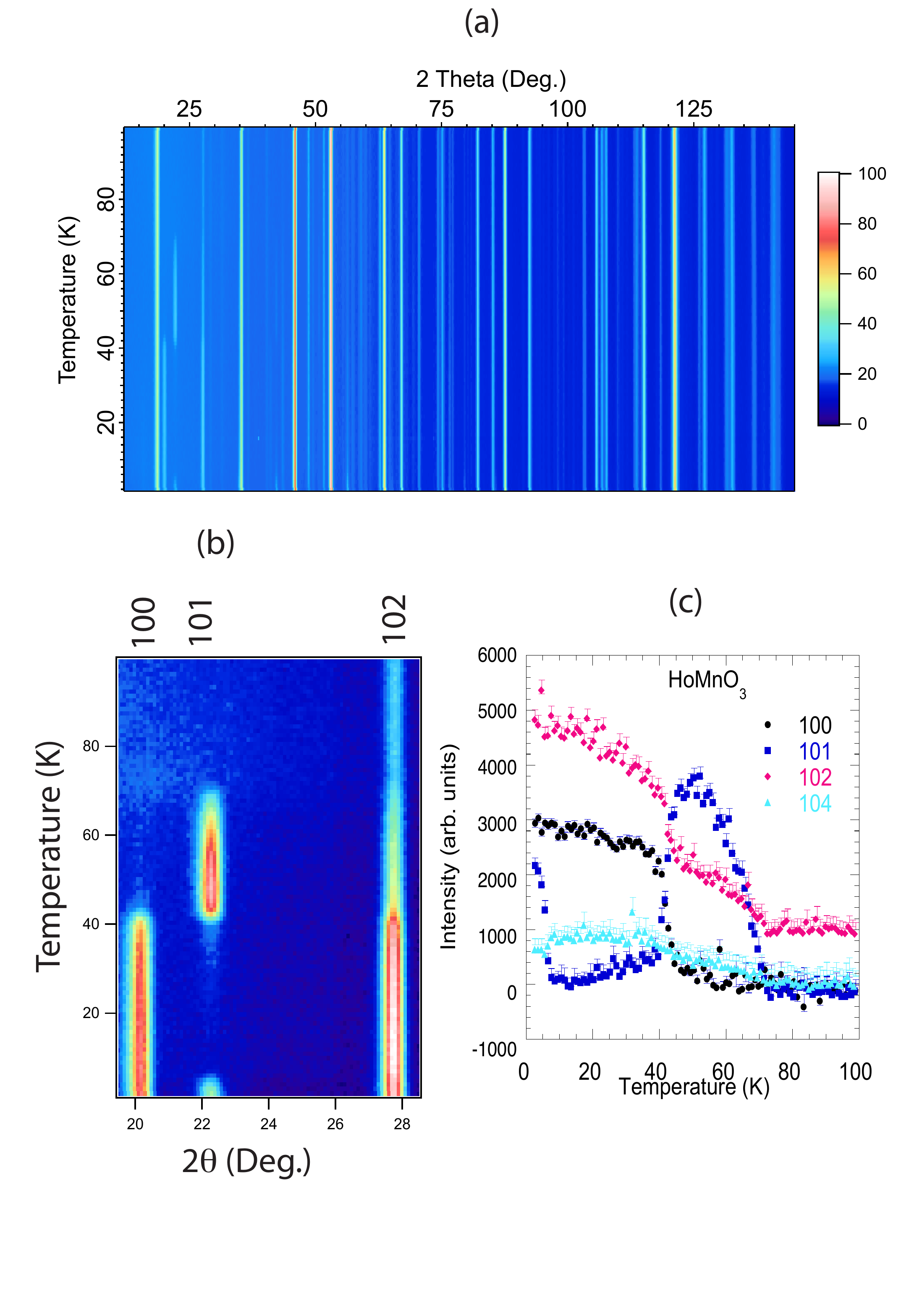}}
\caption {(Color online) (a) Temperature variation of the diffraction intensities of HoMnO$_3$. (b) Temperature variation of a few low angle reflections from HoMnO$_3$ containing magnetic contributions. (c) Temperature variation of the peak intensity of the 100, 101,102 and 104 reflections.}
\label{hmno1}
\end{figure}
\begin{figure}
\resizebox{0.5\textwidth}{!}{\includegraphics{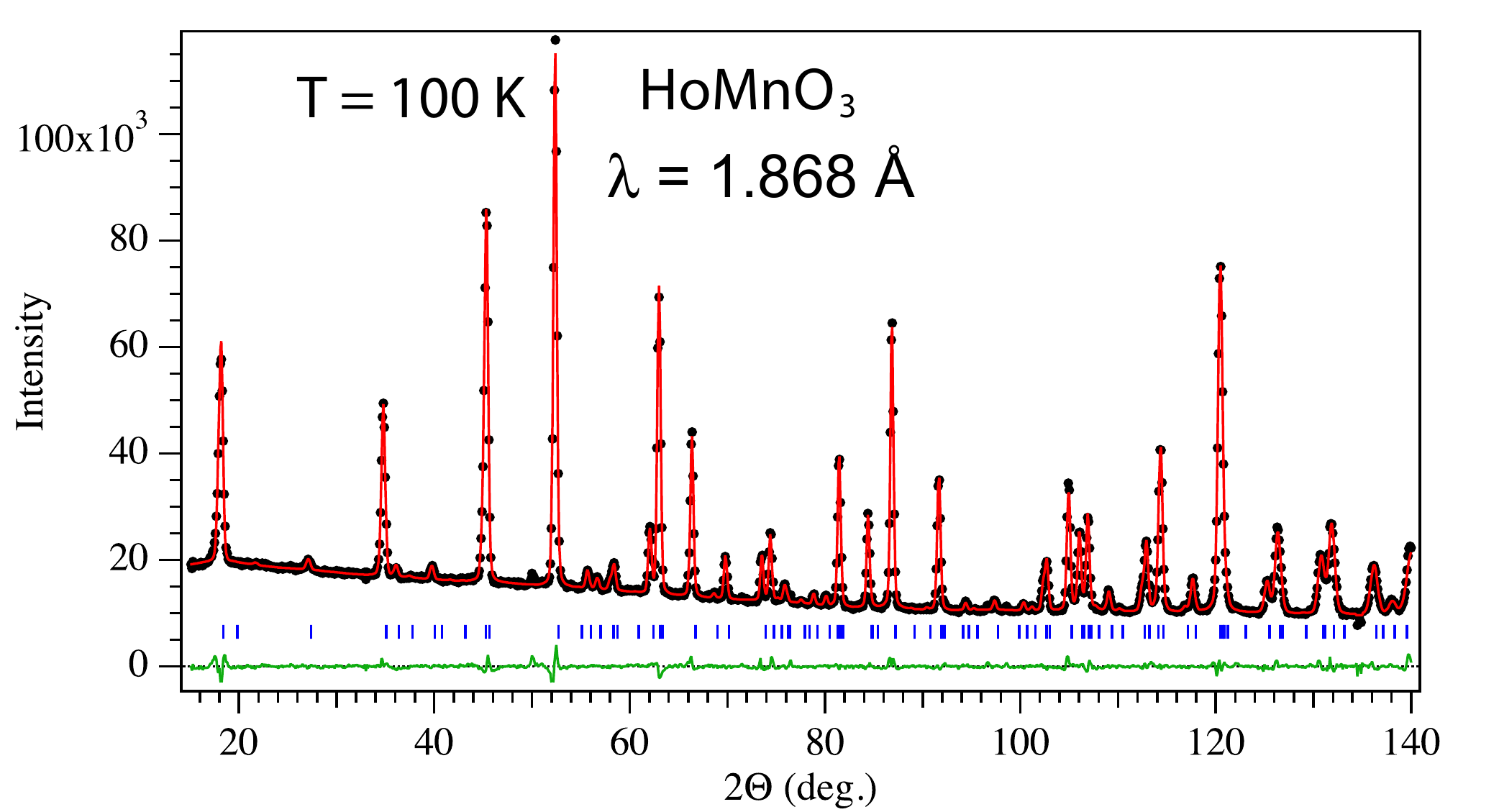}}
\caption {(Color online) Results of the refinement of the hexagonal crystal structure of HoMnO$_3$ at T = 100 K. The tick marks below the diffraction pattern show the calculated positions of the diffraction peaks and the difference between the observed and calculated intensities has been plotted below the tick marks.}
\label{homno_nuc}
\end{figure}

\begin{figure}
\resizebox{0.5\textwidth}{!}{\includegraphics{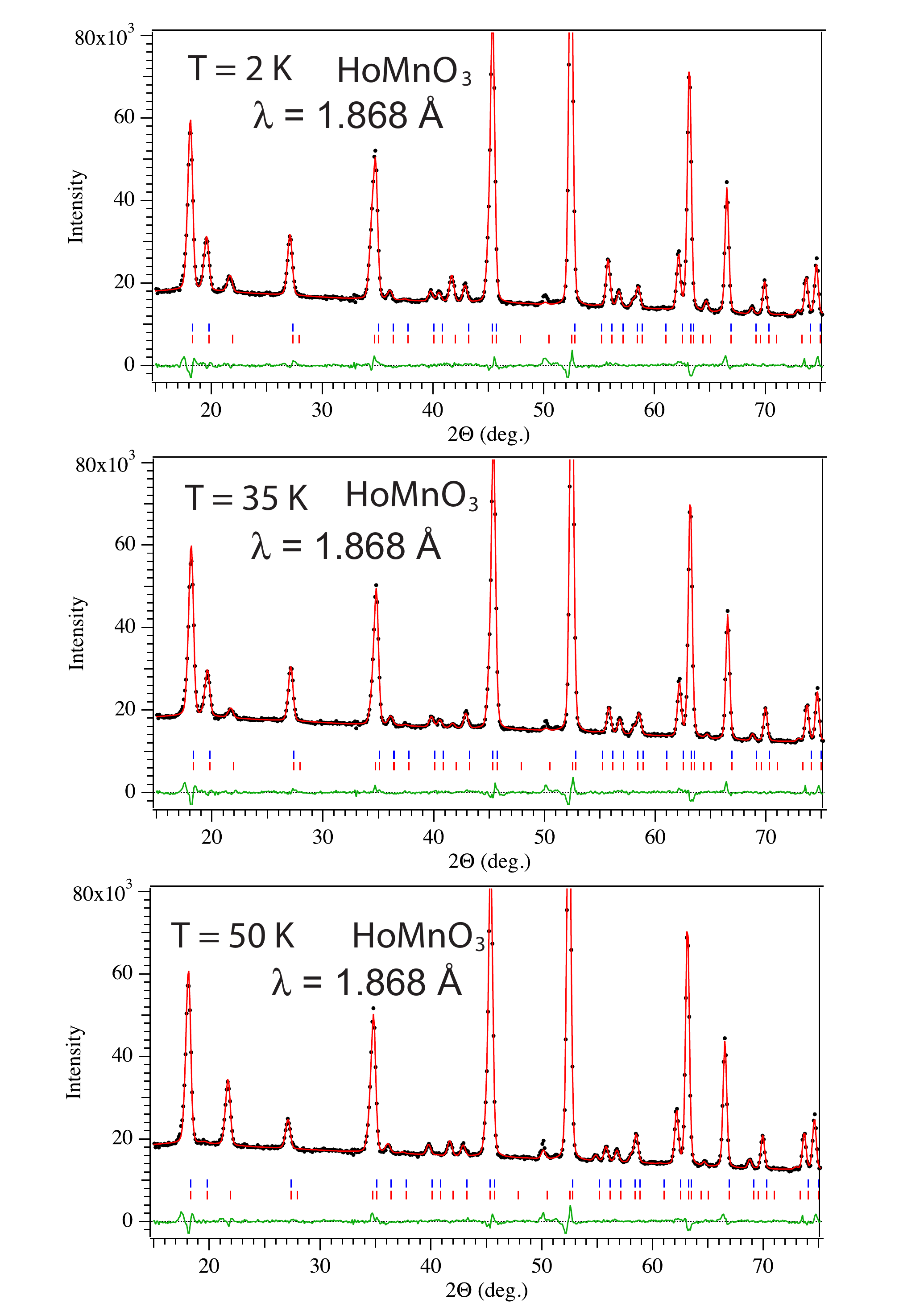}}
\caption {(Color online)  Results of the refinement of the crystal and magnetic structures of HoMnO$_3$ at T = 2, 35 and 50 K.  The upper tick marks below the diffraction pattern show the calculated positions of the nuclear peaks and the lower tick marks show the calculated positions of the magnetic peaks.  The difference between the observed and calculated intensities has been plotted below the tick marks.}
\label{homno_mag}
\end{figure}

\begin{figure}
\resizebox{0.5\textwidth}{!}{\includegraphics{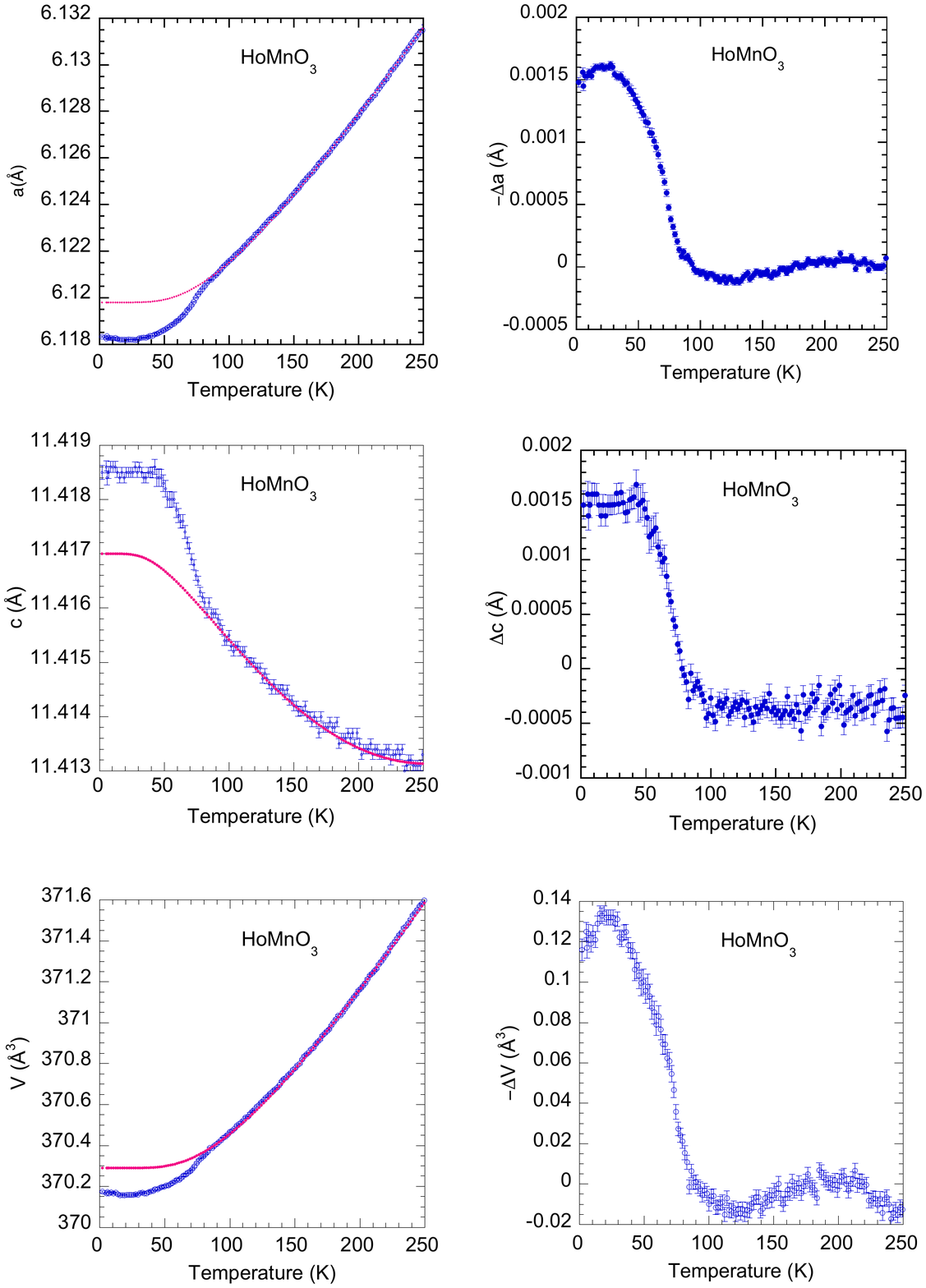}}
\caption {(Color online) Temperature variation of the lattice parameters  $a$, $c$, and the unit cell volume $V$  of HoMnO$_3$ plotted on the left panel. The red curves in these figures represent the lattice parameter and the unit cell volume obtained by fitting the high temperature data by a Einstein-Gr\"uneisen equation and extrapolated to the low temperature. This should give the background for the non-magnetic lattice. On the right panel the temperature dependence of the lattice strains $\Delta a$ and $\Delta V$ obtained by subtracting the non-magnetic background have been plotted.  }
\label{homnolattice}
\end{figure}

\begin{figure}
\resizebox{0.5\textwidth}{!}{\includegraphics{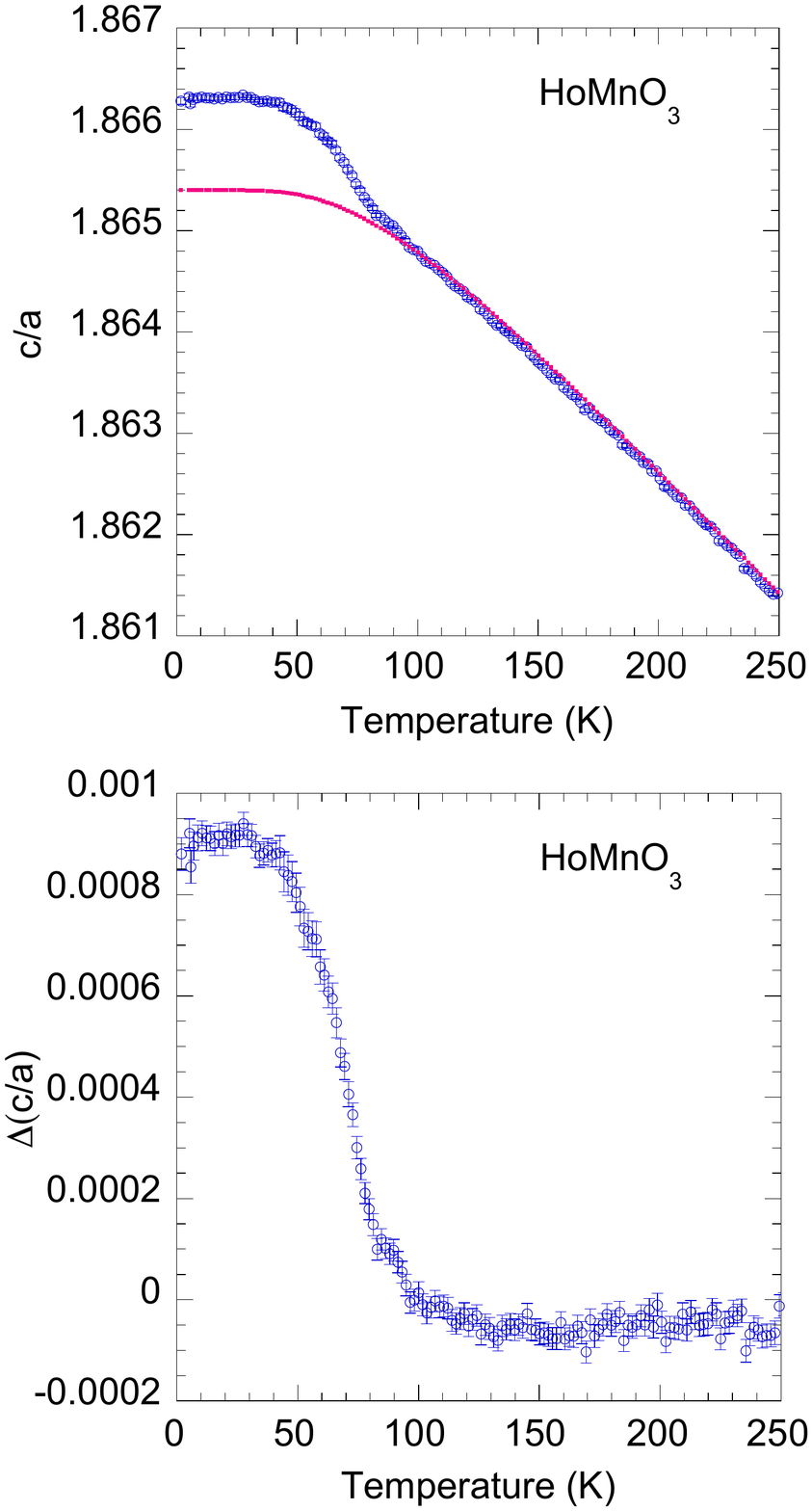}}
\caption {(Color online) Temperature variation of the  $\frac{c}{a}$ ratio  of HoMnO$_3$ plotted on the upper panel. The red curves in these figures represent the  $\frac{c}{a}$ ratio obtained by fitting the high temperature data by the Einstein-Gr\"uneisen function and extrapolated to the low temperature. It is assumed that this gives the temperature variation of  $\frac{c}{a}$ for the non-magnetic lattice. On the bottom panel the temperature dependence of the  $\frac{c}{a}$ ratio obtained by subtracting the non-magnetic background have been plotted.  }
\label{homnoc/a}
\end{figure}

\begin{figure}
\resizebox{0.45\textwidth}{!}{\includegraphics{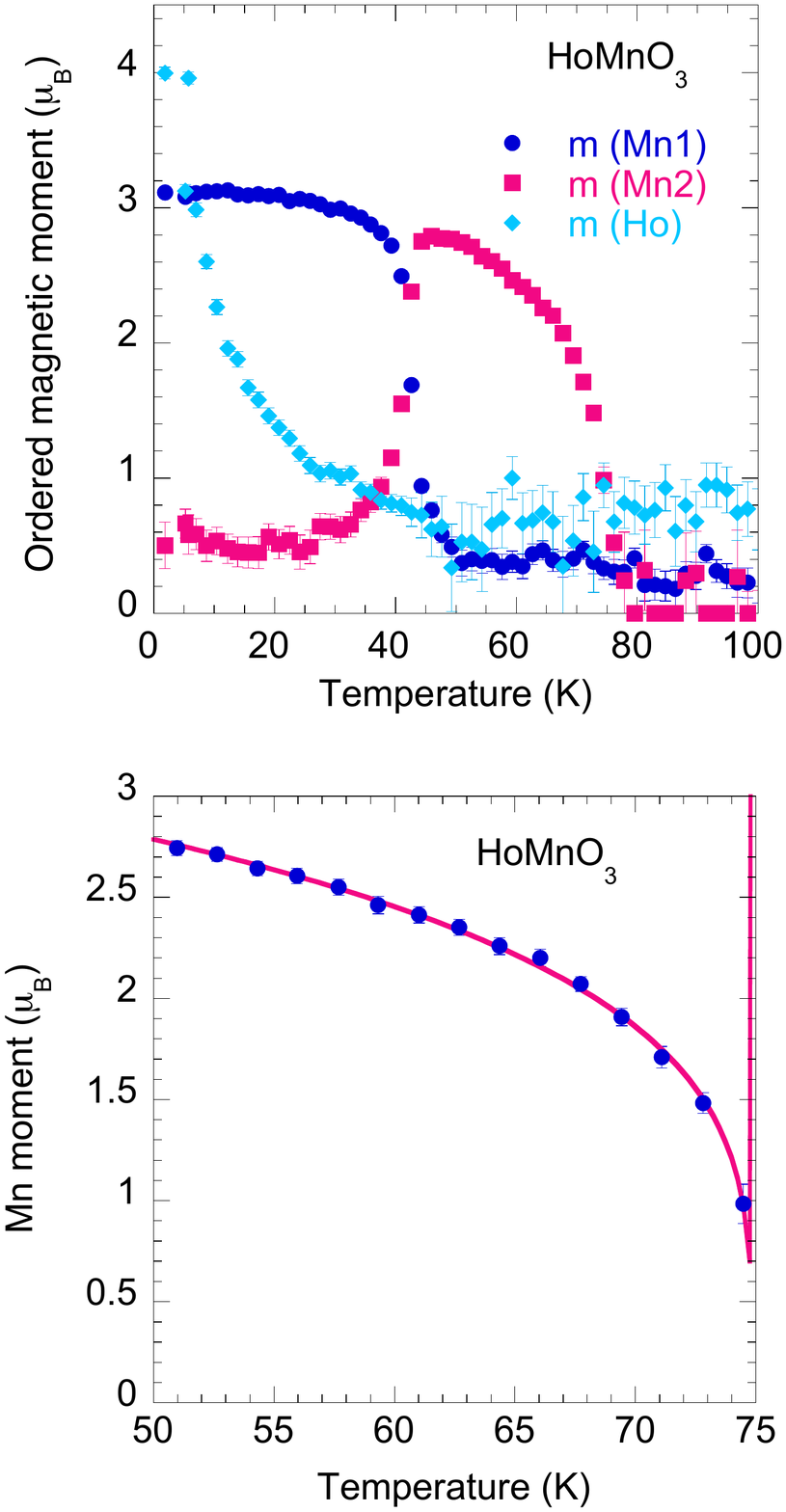}}
\caption {(Color online) (Upper panel)Temperature variation of the ordered magnetic moment m (Mn1) and m (Mn2) of Mn and m (Ho) of Ho ions of HoMnO$_3$ obtained by fitting the measured Bragg intensities of the nuclear and magnetic reflections with the known crystal and magnetic structure models. m (Mn1) is the ordered magnetic moment corresponding to the spin rotated low temperature phase below $T_{SR} \approx 40$ K and m (Mn2) is the ordered magnetic moment of the higher temperature magnetic phase that develops below $T_N \approx 75$ K and undergoes a phase transition below $T_{SR} \approx 40$ K to the spin reoriented low temperature phase. (Lower panel) Power-law fit of the temperature variation of m(Mn2) data close to $T_N$.}
\label{moment}
\end{figure}

HoMnO$_3$ belongs to the family of hexagonal manganites RMnO$_3$ (R = Sc, Y, Er, Ho, Tm, Yb, Lu) that show multiferroic behaviour \cite{huang97}. These hexagonal manganites at paraelectric at high temperatures with the centrosymmetric space group $P6_3/mmc$. Below about 1000 K they undergo a paraelectric-to-ferrielectric transition to a non-centrosymmetric structure with the space group $P6_3cm$. At further lower temperatures of the order of about 100 K the magnetic hexagonal manganites order with a non-collinear antiferromagnetic structure with propagation vector ${\bf k} = 0$. Among these hexagonal manganites YMnO$_3$ and HoMnO$_3$ have been investigated quite intensively \cite{bertaut63,bertaut67,munoz00,munoz01,brown06,brown08,vanaken04}. We have performed  neutron powder diffraction investigation of the temperature dependence of the crystal and magnetic structure of HoMnO$_3$. Here we report the observation of considerable magnetoelastic effects below the N\'eel temperature $T_N \approx 75$ K.

In order to determine quantitatively the magnetoelastic effect or the spontaneous magnetostriction one needs to measure very accurately the temperature variation of the lattice parameters  and the unit cell volume  in small temperature steps. These temperature variations often reveal anomalous behavior around the magnetic ordering temperature. To extract the magnetoelastic effect one needs to know the background temperature variation in the absence of magnetoelastic effect in a fictitious non-magnetic solid which otherwise resembles the compound under study. This is not an easy task and the disagreements between experimental results often arise from the procedure of determination of the background temperature variation of the lattice parameters. One of the commonly employed method is to fit the high temperature data above the ordering temperature by the Debye equation in the Gr\"uneisen approximation \cite{wallace72} or the simpler Einstein equation in the same Gr\"uneisen approximation \cite{chatterji11} and then extrapolating to the lower temperatures to give the background variations in the absence of magnetism. The simpler Einstein-Gr\"uneisen equation \cite{chatterji11} serves as an excellent fit function for not only the volume but also for the individual lattice parameters and we have used it here. 

Neutron diffraction experiments were done on HoMnO$_3$ on the high intensity powder diffractometer D20 \cite{hansen08} of the Institute Laue-Langevin in Grenoble. The $115$ reflection from a Ge monochromator at a high take-off angle of $118^{\circ}$ gave a neutron wavelength of 1.868 {\AA}.  Approximately 5 g of HoMnO$_3$  powder sample was placed inside an $8$ mm diameter vanadium can, which was fixed to the sample stick of a standard $^4$He cryostat. We have measured the diffraction intensities from HoMnO$_3$ as a function of temperature in the range $2 - 300$ K. Fig. \ref{hmno1} (a) shows the temperature variation of the diffraction intensities of HoMnO$_3$. The temperature variation of a few low angle reflections from HoMnO$_3$ containing magnetic contributions is shown in Fig. \ref{hmno1} (b) and the temperature variation of the peak intensities of the 100, 101,102 and 104 reflections is shown in Fig. \ref{hmno1} (c). The temperature variation of the intensities of the low angle reflections from HoMnO$_3$ clearly shows two magnetic phase transitions, one at the N\'eel temperature  $T_N \approx 75$ K and the second spin reorientation transition at about $T_{SR} \approx 40$ K. 

Rietveld refinement \cite{rietveld69} of crystal and magnetic structures against the experimental diffraction data was done by the Fullprof program \cite{rodriguez10}. The refinement results from HoMnO$_3$ at T = 100 K  in the paramagnetic state are shown in Fig. \ref{homno_nuc}. At T = 100 K only the nuclear Bragg peaks were observed and the refinement of the nuclear structure was done by the known crystal structure of the hexagonal manganites \cite{munoz00,munoz01}. Fig. \ref{homno_mag} shows the results of refinement of the crystal and magnetic structures \cite{munoz01,brown06} at T = 2, 35 and 50 K in the three magnetic phases of HoMnO$_3$.The agreement factors and $\chi^2$ values of these refinements are given in Table \ref{agreement}. It is to be noted that magnetic structures of hexagonal manganites including HoMnO$_3$ show homometry, which cannot be distinguished by powder neutron diffraction. Only polarized neutron diffraction on single crystals can resolve the homometric structures \cite{brown06}. 

 \begin{table}[ht]
\caption{Agreement factors for the refinement of crystal and magnetic structures of HoMnO$_3$ at different temperatures. $R_p$ and $R_{wp}$ are the unweighted and weighted agreement or R factors whereas the $R_{exp}$ is the expected R factor corresponding to the statistics of the data and $\chi^2$ has the usual statistical meaning.}
\label{agreement}
\begin{center}
\begin{tabular}{ccccc} \hline \hline
\emph{T(K)} & $R_p$(\%) &$R_{wp}(\%)$& $R_{exp}$(\%) &$\chi^2$\\ \hline
2 & 8.13& 7.11&3.15&5.10\\
35& 8.04& 7.03&3.20&4.82\\
50& 8.53& 7.38&3.22&5.26\\
100& 8.84& 7.35&3.30&4.98\\\hline
\end{tabular}
\end{center}
\end{table}

Figure \ref{homnolattice} shows the temperature variation of the lattice parameters  $a$, $c$, and the unit cell volume $V$  of HoMnO$_3$ plotted on the left panel. The lattice parameter $a$ decreases continuously with decreasing temperature in a very normal way, but close to the N\'eel temperature $T_N \approx 75$ K shows the magnetoelastic or magnetostriction anomaly. The red curve represent the background variation of the $a$ lattice parameter for a nonmagnetic solid obtained by fitting the data in the paramagnetic state by the Einstein-Gr\"uneisen equation explained before. By subtracting the background from the data we determined the lattice strain $\Delta a$ plotted on the right panel of Figure \ref{homnolattice}. The $c$ lattice parameter increases continuously on decreasing temperature down to $T_N$ and then shows abrupt anomalous increase or positive magnetostriction. By subtracting the background variation of the $c$ lattice parameter (red continuous curve) from the data we determined the lattice strain $\Delta c$ plotted on the right panel. Similar plots for the unit cell volume $V$ and and the volume strain $\Delta V$ are shown in Figure \ref{homnolattice} in the left and the right panels, respectively.   Fig. \ref{homnoc/a}  shows in the upper panel the temperature dependence of the $\frac{c}{a}$ ratio. It varies linearly with temperature above about 100 K and below this temperature the $\frac{c}{a}$ shows anomalous behavior due the magnetoelastic effect. However we fitted the  $\frac{c}{a}$ ratio above 100 K by the Einstein-Gr\"uneisen equation and extrapolated to the lower temperature. The fit is shown by the red line. Fig  \ref{homnoc/a}  shows in the lower panel the difference between the observed and the fitted  $\frac{c}{a}$ ratio. It is evident that the temperature variation of  $\frac{c}{a}$ also behave like the other lattice parameters with respect to the magnetoelastic effect. We note that although the magnetic ordering temperature of HoMnO$_3$ is $T_N \approx 75$ K, the magnetoelastic effect starts becoming appreciable already below about $T^* \approx100$ K. This appearance of magnetoelastic effect already about 25 K above $T_N \approx 75$K is due to short-range spin correlations. The spontaneous linear magnetostriction along the a-axis at T = 0 is $\Delta a/a_0=-2.45 \times 10^{-4}$ and that along the c-axis is  $\Delta c/c_0=1.31 \times 10^{-4}$. The spontaneous volume magnetorestriction in HoMnO$_3$ is  $\Delta V/V_0=-3.51 \times 10^{-4}$. We can calculate the  spontaneous volume magnetorestriction in HoMnO$_3$ from $\Delta V/V_0 = 2\Delta a/a_0+\Delta c/c_0 = -3.59 \times 10^{-4}$ which very close to that obtained from the unit cell volume variation $\Delta V/V_0=-3.51 \times 10^{-4}$. This suggests that our method of extracting magnetoelastic effect is reliable or at least consistent. The magnitude of spontaneous magnetostriction in HoMnO$_3$ is relatively large and is comparable to those \cite{chatterji12} determined in the other multiferroic hexagonal manganite YMnO$_3$.

Figure \ref{moment} (a) shows the temperature variation of the ordered magnetic moments $m$  of Mn ion corresponding to the high temperature magnetic phase of HoMnO$_3$ (shown by the red filled squares) and lower temperature spin reoriented phase of HoMnO$_3$ (shown by filled blue circles) obtained by fitting the measured Bragg intensities of the nuclear and magnetic reflections with the known crystal and magnetic structure models. The ordered magnetic moment of the high temperature magnetic phase decreases with increasing temperature above about the spin reorientation transition temperature $T_{SR} \approx 40$ K in the usual way and becomes zero at $T_N \approx 75$ K. Below $T_{SR} \approx 40$ K the moment of this phase decreases and the magnetic moment of the spin reoriented low temperature phase increases abruptly and becomes saturated to about $3.1 \mu_B$ at T = 2 K. The Ho magnetic moment is polarized at all temperatures below $T_N \approx 75$ K but below about 50 K the Ho moment increases with decreasing temperature and becomes about $4\mu_B$ at T = 2 K. The temperature variation of the ordered magnetic moment of the Ho sublattice resembles the Brillouin function and indicates strongly the absence of any magnetic phase transition \cite{brown08} in HoMnO$_3$ at lower temperatures. Figure \ref{moment} (b) shows the power-law fit of the magnetic moment of Mn in the higher temperature phase of HoMnO$_3$ below   $T_N \approx 75$ K. We have included the data points in the temperature range 50-74 K. The least squares fit of these data gave the power-law exponent $\beta = 0.247 \pm 0.005$ and $T_N = 74.84 \pm 0.05$ K. Since there are no data close to $T_N$ the the power-law exponent obtained cannot be associated really with the critical exponent. However the determined N\'eel temperature $T_N = 74.84 \pm 0.05$ K from the fit is dependable within the experimental resolution. It is to be noted that the magnetoelastic effects shown in Fig. \ref{homnolattice} in the lattice parameters are not sensitive to the spin-reorientation transition at $T_{SR} \approx 40$ K. It is not quite sure whether the ordered magnetic moment of the Ho sublattice at lower temperature influence the magnetoelastic effect substantially.

The temperature variations of the lattice strains determined in HoMnO$_3$ do not look very smooth and do not become zero above $T_N \approx 75$ K. This is partly due to the problem of determining the background temperature variation of the lattice parameters and unit cell volume by the Einstein-Gr\"uneisen function and also may be due to multiple magnetic phase transitions and the polarization of the Ho magnetic moments. We could not therefore relate quantitatively the lattice strain and the magnetic order parameter as we did in YMnO$_3$ and other simple antiferromagnetic systems \cite{chatterji12,chatterji10,chatterji10a}. We checked carefully whether we could get any information about the shifts in atomic coordinates in the unit cell below the magnetic ordering temperature or the internal magnetoelastic effect. Although the agreement factors of the refinement of the crystal and magnetic structures against the data were reasonably good (see Table \ref{agreement}), the resulting positional parameters do not show within the refinement accuracy any anomalous shifts in atomic coordinates below $T_N$. 

In conclusion we have measured accurately neutron diffraction intensities of HoMnO$_3$ powder sample and determined the magnetostriction below the magnetic ordering temperature. We have also determined the magnetic phase transitions and the temperature evolution of the magnetic moments of Mn and Ho ions.


\begin{thebibliography}{99}
\bibitem{becker39}R. Becker and W. D\"oring, \emph{Ferromagnetismus}(Verlag Julius Sringer, Berlin, 1939)
\bibitem{callen63}E.R. Callen and H.B. Callen, Phys. Rev. B {\bf 129}, 578 (1963).
\bibitem{callen65}E.R. Callen and H.B. Callen, Phys. Rev. B {\bf 139}, A455 (1963).
\bibitem{lee55}E.W. Lee, Rep. Prog. Phys. {\bf 18}, 184 (1955).
\bibitem{clark80}A.E. Clark, \emph{Ferromagnetic Materials}, ed. E.P. Wohlfarth, North-Holland Publishing Company (1980).
\bibitem{andreev95}A.V. Andreev, \emph{Handbook of Magnetic Materials}, ed. K.H. Buschow, Elsevier, vol. 1, p. 531 (1995).
\bibitem{lindbaum02}A. Lindbaum and M. Rotter, \emph{Handbook of Magnetic Materials}, ed. K.H. Buschow, Elsevier, vol. 14, p. 307 (2002).
\bibitem{doerr05}M. Doerr, M. Rotter and A. Lindbaum, Advences in Physics, {\bf 54}, 1 (2005).
\bibitem{huang97}Z.J. Huang, Y. Cao, Y.Y. Sun, Y.Y. Xue and C.W. Chu, Phys. Rev.B {\bf 56}, 2623 (1997).
\bibitem{bertaut63}E.F. Bertaut and M. Mercier, Phys. Lett. {\bf 5}, 27 (1963).
\bibitem{bertaut67}E. Bertat, R.Pauthenet, and M. Mercier, Phys. Lett. {\bf 18}, 13 (1967).
\bibitem{munoz00}A. Mun\~oz, J.A. Alonso, M.J. Mart\'inez-Lope, M.T. Cas\'ais, J.L. Mart\'inez, and M.T. Fern\'andez-D\'iaz, Phys. Rev. B {\bf 62}, 9498 (2000).
\bibitem{munoz01}A. Mun\~oz, J.A. Alonso, M.J. Mart\'inez-Lope, M.T. Cas\'ais, J.L. Mart\'inez, and M.T. Fern\'andez-D\'iaz, Chem. Mater.  {\bf 13}, 1497 (2001).
\bibitem{brown06}P.J. Brown and T. Chatterji, J. Phys: Condens. Matter {\bf 18}, 10085 (2006).
\bibitem{brown08}P.J. Brown and T. Chatterji, Phys. Rev. B {\bf 77}, 104407 (2008).
\bibitem{vanaken04}B. B. van Aken, T.T.M. Palstra, A. Filippetti and N.A. Spaldin, Nature Materials {\bf 3}, 164 (2004).
\bibitem{wallace72}W.C. Wallace, \emph{Thermodynamics of Crystals}, John Wiley \& Sons, NewYork (1972).
\bibitem{chatterji11}T. Chatterji and T.C. Hansen, J. Phys.: Condens. Matter {\bf 23}, 276007 (2011).
\bibitem{hansen08}T.C. Hansen, P.F. Henry, H.E. Fischer, J. Torregrossa and P. Convert, Meas. Sci. Technol. {\bf 19}, 034001 (2008).
\bibitem{rietveld69}H.M. Rietveld, J. Appl. Cryst. {\bf 2}, 65 (1969).
\bibitem{rodriguez10}J. Rodriguez-Carvajal, FULLPROF, a Rietveld and pattern matching and analysis program version 2010, LLB, CEA-CNRS, France [http://www.ill.eu/sites/fullprof/]
\bibitem{chatterji12}T. Chatterji, B. Ouladdiaf, P.F. Henry and D. Bhattacharya, J. Phys.: Condensed Matter {\bf 24}, 336003 (2012).
\bibitem{chatterji10}T. Chatterji, G.N. Iles, B. Ouladdiaf and T.C. Hansen, J. Phys.: Condensed Matter {\bf 22}, 316001 (2010).
\bibitem{chatterji10a}T. Chatterji, B. Ouladdiaf and T.C. Hansen, J. Phys.: Condensed Matter {\bf 22}, 096001 (2010).


\end{thebibliography}
\end{document}